\documentclass[aps,pra,twocolumn,showpacs,letterpaper,tighten,float,superscriptaddress,nofootinbib,reprint]{revtex4-1}

\usepackage[utf8]{inputenc}
\usepackage[fleqn]{amsmath}
\usepackage[mathscr]{euscript}
\usepackage{amsfonts}
\usepackage{amssymb}
\usepackage{array}
\usepackage[T1]{fontenc}
\usepackage[margin=1in]{geometry}
\usepackage{lmodern}
\usepackage{tabularx}
\usepackage{fancyhdr}
\usepackage{graphicx}
\graphicspath{{images/}}
\usepackage[utf8]{inputenc}
\usepackage[english]{babel}
\usepackage{amsthm}
\usepackage{braket}
\usepackage{tikz}
\usepackage{subfig, float}
\usepackage{algorithm}
\usepackage{algpseudocode}

\newtheorem{theorem}{Theorem}

\newcommand{\hlf}{\frac{1}{2}}

\newcommand{\tr}{\text{Tr}}

\newcommand{\mc}{\mathcal}

\newcommand{\mb}{\mathbb}
\newcommand{\id}{\text{id}}

\newcommand{\im}{\text{Im }}

\newcommand{\prob}{\text{Prob}}

\begin{document}
\title{Thermal Stability of Dynamical Phase Transitions in Higher Dimensional Stabilizer Codes}
\author{Albert T. Schmitz}
\email{albert.schmitz@colorado.edu}
\affiliation{
Department of Physics and Center for Theory of Quantum Matter,
University of Colorado, Boulder, Colorado 80309, USA}


\begin{abstract}
For all of the interest in dynamical phase transitions (DPT), it is still not clear the meaning or prevalence of these features in higher dimensional models. In this paper, we consider DPTs for stabilizer code models and quantum quenches between these models in higher dimensions, particularly $d=2,3$. We find that for many stabilizer codes, there exists a robust DPT to thermal noise which indicates the resilience of information stored in the ground space in the context of quantum error correction. That is, the critical temperature at which the DPT is lost corresponds to the theoretical upper-bound on the decoding rate for the code. We also discuss a generalization of the Wegner duality and how it can be used to characterize DPTs and other thermal properties.  
\end{abstract}

\maketitle

\section{Introduction}

The study of quantum systems has historically been focused on static features such as the properties of ground states and thermal equilibrium. This has been the case in part due to the experimental limitations on studying non-equilibrium states and the overall success of non-interacting fermion theory at zero temperature in explaining material properties. However, the community focus has shifted in recent years due to advances in experimental systems such as cold-atoms, quantum dots and superconducting circuits which are capable of manipulating the dynamics of fundamentally quantum systems, and the use of these systems for quantum information processing. Further theoretical advances have also shifted attention to dynamical systems with the exploration of phenomena such as many-body localization, quantum chaos, Floquet states, quantum scars, and many others. But just as with static and equilibrium properties, we look to find universal features in dynamical systems which are robust to noise and small-scale variation in the underlying models \cite{Gambassi2012}. One candidate for such a characterization is the existence and form of {\it dynamical phase transitions} (DPT) \cite{Heyl2013}.  These are defined as non-analytic behavior in the dynamical free energy density,
\begin{align}\label{eq:freeE}
F(t)= - \lim_{N\to \infty} \frac{1}{2N} \ln|\tr(U(t))|^2,
\end{align}
where $U(t)$ is the unitary operator describing the time-evolution of the system and $N$ is the number of degrees of freedom.\footnote{As $\tr(U(t))$ can in principle be zero for non-universal/uninteresting reasons, we also require that a DPT be finite in the free energy density and so the non-analytic behavior occurs for the derivatives of $F(t)$.} In practice, such an object can be difficult to measure, so a DPT is also extended to the analogous free energy density for a {\it quantum quench} or {\it Loschmidt echo}, where $\tr(U(t))$ in Eq. \eqref{eq:freeE} is replaced by $\braket{\psi|U(t)|\psi}$ for a state $\ket{\psi}$ with support on a  significant number of energy eigenstates. Practically, such a quantity is measured by letting the system settle into thermal equilibrium at $T\approx 0$  for one Hamiltonian, $H'$, with ground state $\ket{\psi}$, at which point the system is ``quenched'' with another Hamiltonian, $H$, representing the time evolution $U(t)= \exp(-it H)$. After a time $t$, the system is then measured in a basis for which $\ket{\psi}$ is an eigenstate. DPTs have already been observed in experimental cold atom systems \cite{Jurcevic2017, Flaschner2018}, spurring a great deal of interest in their theoretical study.

Theoretical study has primarily focused on one-dimensional systems such as the $d=1$ traverse-field Ising model \cite{Heyl2013, Heyl2015}, long-range Ising chain model \cite{Zunkovic2018, Halimeh2017, Halimeh2017a, Zauner2017}, and random-field Ising model \cite{Gurarie2019}, to name a few \cite{Heyl2014, Karrasch2013, Kriel2014, Andraschko2014, Hickey2014, Vajna2014}. Some progress has been made for two-dimensional models such as the $d=2$ transverse-field Ising model \cite{Hashizume2018, Hashizume2019} and the ``extended'' toric code model \cite{Srivastav2019} where on-site fields are added to the $d=2$ toric code \cite{Kitaev2003} in such a way that the Hamiltonian can be mapped to several copies of the $d=1$ traverse-field Ising model. The only works we are aware of for higher dimensional models is Ref. \cite{Canovi2014} which studies higher-dimensional fermionic models using dynamical mean-field theory and work on the infinite-dimensional quantum Ising model \cite{Homrighausen2017, Lang2018, Lang2018a}. It has also been established that there is some connection between equilibrium phase transitions and DPTs \cite{Heyl2013, Heyl2014, Kriel2014,Karrasch2013}, though the connection is not rigid \cite{Vajna2014}. The study of the robustness of this phenomenon has also been extended to thermal states via a quench of the Gibb-state density matrix \cite{Heyl2017, Heyl2017a,Bhattacharya2017, Lang2018, Abeling2016}.  The most relevant of these results for this paper are found in Refs. \cite{Bhattacharya2017, Heyl2017} which find a persistence in the DPT related to the topological properties of free-fermion systems.

In this paper, we look to study DPTs and their thermal stability for a collection of integrable models which can be described as Pauli {\it stabilizer codes} \cite{Gottesman1997}. A stabilizer code is an example of a quantum error-correcting code (QECC) which is described by a set of mutually commuting, non-trivial, Pauli operators, $S$. Information is then stored in the subspace of the Hilbert space which is stabilized by these operators. See Ref. \cite{Terhal2015} for a review. If the stabilizers are of either X- or Z-type, but not both, then the stabilizer code is generally referred to as a Calderbank, Shor and Steane (CSS) code. Such codes also serve as exactly-solvable models for conventional topological phases such as the $d=2$ toric code \cite{Dennis2002} or more exotic phases such as {\it fracton topological phases}  \cite{ Chamon2005, Castelnovo2012, Haah2011, Haah2013, HaahThesis, Vijay2016, Schmitz2019c}. Fracton models--generally in three-dimensions--are defined as containing quasi-particles with geometrically limited or absent local hopping due to a sub-extensive number of conservation laws \cite{Pretko2017a, Pretko2017b, Schmitz2019a, Nussinov2009a}. See Refs. \cite{Nandkishore2018, Pretko2020} for a review. We show that DPTs do exist for most stabilizer codes including fractonic stabilizer codes in three-dimensions. However if the code does not have a thermal phase transition, then their nature and form (in the exact stabilizer limit) cannot be distinguished from those found in the $d=1$ Ising model. Moreover for CSS codes, when the Loschmidt echo initial state is a thermal product eigenstate of either all X or Z Pauli operators, the DPT is robust to thermal noise up to a certain critical temperature which corresponds to the theoretical upper bound on the decoding rate of the underlying QECC, i.e. it represents the robustness of information stored in the code space. We show this by using a generalization of the Wegner duality \cite{Wegner1971} and an abstraction of stabilizer codes to a {\it linear gauge structure}\cite{Schmitz2019a}. This allows us to make several generic statements that can be applied to any stabilizer code.  

\subsection{Overview of Results}

The primary result of this paper is the following statement: for many stabilizer codes (specifically CSS type), any DPT in a quantum quench (Loschmidt echo) between a trivial product ground state and driven by the stabilizer code time evolution is robust to thermal noise, if there exists an ordered phase for the transpose model as defined below.\footnote{The name ``transpose'' is not common in the literature and is justified by the definition below. The model is often referred to as a dual or gauge dual \cite{Vijay2016, Williamson2016, Kubica2018}, but we do not use this terminology to avoid confusion with the generalized Wegner dual which we also discuss below.} The critical temperature at which the DPT is lost corresponds to the critical temperature of the transpose model which, as argued in Ref. \cite{Dennis2002} and reviewed below, is an indication of the robustness of information protection in the ground space. This is irrespective of dimension, however a thermal phase transition in the transpose model generally only exists in dimensions $d>1$.

Some of the secondary results which lead up to this  conclusion are:

\begin{itemize}
\item The thermal partition function for a stabilizer code at low temperature is (Wegner) dual to a high temperature model of its constraints, where a constraint is any product of stabilizers equal to the identity \cite{Weinstein2019}.

\item The thermal model for the transpose of the stabilizer code is self-dual (in the above sense), and a phase transition in this model represents the decoding rate of the stabilizer code, i.e. characterizes how well information is protected in the ground space.

\item The dynamical partition function also has a dual representation in terms of constraints which can be used to characterize DPTs.

\item Stabilizer codes with only extensive or sub-extensive constraints have no thermal phase transition and a DPT which is equivalent to the DPT of the $d=1$ Ising model; this includes most fracton stabilizer codes. 

\item The quantum quench (or Loschmidt echo) partition function between two stabilizer codes can also be described with a constraint-like dual.

\end{itemize}

Before discussing these results in detail, we elaborate on the abstraction of stabilizer codes to a linear gauge structure \cite{Schmitz2019a} as it is necessary for understanding the context and arguments for these results. Likewise, it gives a sense of how conservation laws among excitations of the stabilizer Hamiltonian--as dictated by constraints--determines the structure of both the thermal and dynamical partition functions, thus the existence of DPTs and their thermal stability.

The remainder of the paper is structured as follows: throughout the rest of the paper, we use the $d=2$ toric code as an example to demonstrate the concepts and results as they are introduced. As such, we begin by giving a basic introduction to the $d=2$ toric code. We then give an introduction to linear gauge structures as used for stabilizer codes, and the most important consequence of this, the {\it Braiding Law for Excitations} or BrLE rules. This leads naturally to a discussion of the generalized Wegner duality which when combined with the BrLE rules implies the Wegner dual for any stabilizer code is a model of its constraints. We then apply this duality to the real-time partition function from which we can imply no thermal phase transition and a DPT equivalent to the $d=1$ transverse-field Ising model for stabilizer codes with only (sub) extensive constraints. We show this explicitly for the simplest fracton model, the $d=2$ Ising plaquette model. From there, we move on to stabilizer quantum quenches by introducing some important derived linear gauge structures, including the transpose. We discuss how the transpose model generally contains a thermal phase transition which is indicative of the robustness of information storage of the QECC. This allows us to apply the BrLE rules and generalized Wegner dual to a quantum quench between two stabilizer codes which are described by these derived structures. This leads to our primary result regarding the thermal robustness of DPTs when applied to the quench of a thermal Gibbs state.  

\section{Generic Properties of Stabilizer Code Partition Functions}

\begin{figure}[t]

\centering

\includegraphics[scale=.4]{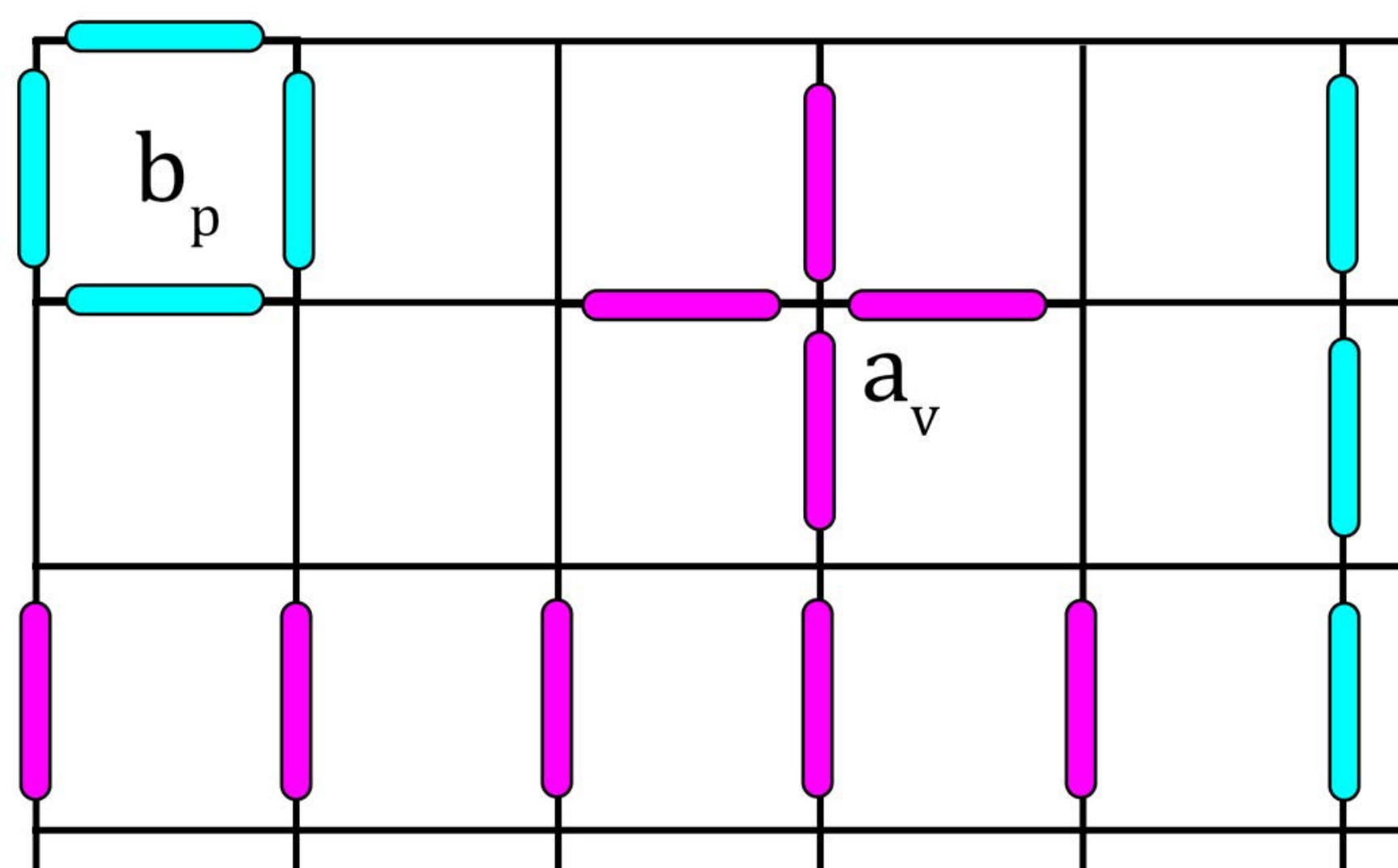}

\caption{Depiction of the operators defining the $d=2$ toric code as well as the string logical operators. Magenta represents X-type operator support and cyan Z-type operator support.}\label{fig:tcdef}

\end{figure}

\subsection{Demonstration Example: $d=2$ Toric Code}

To demonstrate our general results as they are introduced, we use the $d=2$ toric code as first discussed in Ref. \cite{Kitaev2003} as an example throughout the next two sections. This is a CSS stabilizer code with a Hilbert space of $N=2L^2$ qubits arranged on the $d=2$ lattice (2-torus) such that each edge is associated to one qubit. The Hamiltonian terms are then given by,
\begin{align}
S_{TC_2} = \{a_v, b_p: v \text{ vertices and } p \text{ plaquettes}\}.
\end{align}
$b_p= \prod_{i \in p} z_i$, where $i$ indexes the edge qubits about the plaquette $p$ and $a_v = \prod_{i @ v} x_i$, where $i$ indexes the edge qubits coordinated to the vertex $v$.  The Hamiltonian for this model is then given by 
\begin{align}\label{eq:tcham}
H_{TC_2}= \hlf \sum_v (1- a_v) + \hlf \sum_p (1-b_p),
\end{align}
i.e. is the sum over the projection operators onto the $-1$ eigenstates of the stabilizers.

Where needed, we also use $S_X= \{x_i\}$ as another stabilizer code with an analogous Hamiltonian to Eq. \eqref{eq:tcham} as the sum of projection operators onto the $\ket{-}$ eigenstate for each qubit.

\subsection{Stabilizer Codes as Examples of Linear Gauge Structures}\label{sec:LGS}
 
Let $S$ be a set of mutually-commuting Pauli operators on $N$ qubits. We can generically represent all products of these stabilizer by the set $\mc A = \wp(S) \simeq \{0,1\}^{|S|}$, i.e. the power set of stabilizers which is equivalent to a binary vector space with as many bits as there are stabilizers.\footnote{Addition for the power set is given by the symmetric difference of sets, i.e. $A+B= A\cup B - A\cap B$.} For $\text{TC}_2$, it is as if we attach a classical Ising spin to each vertex and each plaquette, and all patterns of up and down spins collectively form $\mc A_{TC_2}$. For any $A\in \mc A$, we use $A_s$ as the bit representing the $s^{th}$ stabilizer and $\hat A_s \simeq \{s\}$ as the unit vector representing stabilizer $s \in S$. We also equip this vector space with a non-degenerate binary two-form for all $A,B \in \mc A$,
\begin{align}
\omega(A, B)= |A \cap B| \mod 2 = \sum_{s\in S} A_s B_s,
\end{align}
where $|\,|$ is the number of elements in the set. For the binary representation, this is equivalent to the binary dot product. To relate these ``virtual'' spins to the actual operators they represent, we consider the map which takes a set of stabilizers (or binary vector representing it) to an operator i.e. for all $A\in \mc A$,
\begin{align}
\phi(A) = \prod_{s \in A} s= \prod_{s\in S} s^{\omega(A, \hat A_s)}.
\end{align}

To make the co-domain of this map precise, we consider the set of all products of single-qubit Pauli operators, $\mc F$,  modding out any phase of $\pm1, \pm i$ i.e. $\mc F=\braket{x_i, z_i}/ U(1)$.  By modding out the phase, $\mc F$ can be treated as a binary vector space of dimension $2N$ which is spanned by single-qubit $X$ and $Z$ Pauli operators. In this vector space, addition is given by the product of the corresponding operators, scalar multiplication is given by the power of that operator, and the zero element is the identity. However without the phase, we need some encoding of the commutation relations. As all Pauli operators either commute or anti-commute, we define another non-degenerate binary two-form $\lambda$ defined for all $f,g \in \mc F$
\begin{align}\label{eq:lamdef1}
\lambda(f,g)= \begin{cases}
0 & \text{ if $f$ and $g$ commute,}\\
1 & \text{otherwise.}
\end{cases}
\end{align}
So the co-domain of $\phi$ is $\mc F$ and $\im \phi$ is the set of all stabilizer products. This function is also linear with respects these two binary vector spaces. For $\text{TC}_2$, $\im \phi_{TC_2}$ represents all trivially-contactable closed strings of $Z$ operators and closed dual-lattice strings of $X$ operators. The fact that all stabilizers are mutually commuting is now encoded in the relation $\lambda(\phi(A), \phi(B))= 0$, for all $A,B\in \mc A$. 

The corresponding Hamiltonian for the system is given by,
\begin{align}\label{eq:stabham}
H_{S} = \hlf \sum_{s \in S}E_s (1-s)=  \hlf \sum_{s \in S}E_s (1-\phi(\hat{A}_s)),
\end{align}
where $E_s>0$ and the phase of $s =\phi(\hat{A}_s) \in S$ is chosen such that it is Hermitian. The ground space of this Hamiltonian is then the mutual $+1$ eigenstates of all the stabilizers. This is also the code space where logical information is stored in the context of QECC\cite{Terhal2015}. To describe the excitations of this Hamiltonian, we consider acting on a ground state with any Pauli operator. This excited state is a $-1$ eigenstate of all stabilizers which anti-commute with that operator. So we encode these excitation patterns for any Pauli operator $f \in \mc F$ with another linear map,
\begin{align}
\psi(f)=& \{ s\in S: \lambda(s, f)=1\} \nonumber \\
\simeq& \sum_i \lambda(\phi(\hat {A}_s), f) \hat{A}_s,
\end{align}
Which maps into $\mc A$. So $\im \psi$ represents all realizable excitation configurations. $\im \psi_{TC_2}$ is given by all configurations of the vertex and plaquette virtual spins which have a even number of down spins (1 bits) due the the $\mb Z_2$ charge conservation of this model. 

With this, we can reduce the Hamiltonian to a classical energy functional on these virtual spins such that for all $J\in \im \psi$,
\begin{align}
E_S(J)= \sum_{s\in S} E_s J_s = \sum_{s\in S} E_s \omega(J,\hat A_s).
\end{align}
The projection operator onto eigenstates with energy $E_S(J)$ is given by
\begin{align}
p_S^{(J)}= \frac{1}{|\im \phi|}\sum_{A \in \mc A} (-1)^{\omega(A, J)} \phi(A).
\end{align}

The pieces, $\phi, \psi, \omega$ and $\lambda$ are all connected as they satisfy the conditions of a linear gauge structure as introduced in Ref. \cite{Schmitz2019a}.  In general, a linear gauge structure consists of two vector space (for our purposes, these spaces are always binary spaces i.e. the field is $\mb F_2$), each equipped with a two-form. The first space, $(\mc A, \omega)$, is referred to as the {\it potential space} and the second, $(\mc F, \lambda)$, is referred to as the {\it field space}.  Along with these spaces, we also have two linear maps $\phi:\mc A \to \mc F$ and $\psi:\mc F \to \mc A$, so that in total, the gauge structure is defined as GS$=\left((\mc A, \omega, \phi),(\mc F, \lambda, \psi)\right)$ which collectively satisfies {\it the braiding relation}\footnote{For an explanation of this name, see Ref. \cite{Schmitz2019a}},
\begin{align}\label{eq:stabgauge}
\lambda(\phi(A), f) = \omega(A, \psi(f)).
\end{align}
That is, these functions are generalized adjoints of each other with respects to these two-forms i.e. $\phi= \psi^\dagger$. 

The most important consequence of a linear gauge structure is the so-called {\it Braiding Law for Excitations} (BrLE rules), given by
 \begin{subequations}\label{eq:conss}
\begin{align}
(\ker \phi)^{\perp_\omega} =& \im \psi, \label{eq:cons}\\
(\im \phi)^{\perp_\lambda} =& \ker \psi \label{eq:consalt},
\end{align}
\end{subequations}
where $(\ker \phi)^{\perp_\omega}$ refers to the set of all $A\in \mc A$ such that $\omega(A,B)=0$ for all $B\in \ker \phi$ and likewise for $(\im \phi)^{\perp_\lambda}$ using $\lambda$. These results are proven in general and for stabilizer codes in Ref. \cite{Schmitz2019a}. Thus we can relate all realizable excitation patterns to members of $\ker \phi$ which is referred to as the {\it constraint space}. The constraint space represents all products of stabilizers which are proportional to the identity, i.e. for all $C\in \ker \phi$, $\phi(C)=0 \simeq \id$. So Eq. \eqref{eq:cons} represents a $\mb Z_2$ charge conservation law as it says {\it any excitation configuration must overlap with all constraints an even number of times}. For example, $\ker \phi_{\text{TC}_2}$ is generated by two independent constraints containing all stabilizers of a given type, 
\begin{subequations}
\begin{align}
\phi(\{a_v\})=\prod_v a_v = \id,\\ 
\phi(\{b_p\})= \prod_p b_p = \id.
\end{align}
\end{subequations}
So for each stabilizer type or ``sector'', only an even number of excitations can overlap with this global constraint an even number of times, thus enforcing global $\mb Z_2$ charge conservation. However, if we alter the boundary conditions and trivialize the topology such that these constraints are lost, then we are allowed to have an odd number of excitations by extending an open string operator to annihilate an excitation at the boundary. A constraint which is lost once we change the topology of the underlying system is referred to as a {\it topological constraint}, and all other constraints are referred to as {\it local constraints} as they are often generated locally (see Ref. \cite{Schmitz2019a} for a rigorous distinction).

The second BrLE rule, Eq. \eqref{eq:consalt} may appear more trivial. It states that the only operators which commute with all stabilizers--which includes $\im \phi$--are those in $\ker \psi$. $\ker \psi /\im \phi$ then represents all operators which commute with the stabilizers, but are not products of stabilizers themselves. These are referred to as {\it logical operators} as they form the Pauli algebra for information encoded in the ground space in the context of QECC\cite{Terhal2015}. For $\text{TC}_2$, these are the deformable Wilson loops or string operators wrapping the 2-torus.  

When the stabilizer model is classical, i.e. all operators trivially commute, it is clear all the above gauge structure results still apply. However, we can avoid any notion of ``quantum-ness'' and represent these models in a purely classical form. This is done by replacing $\mc F$ with the space of physical Ising spins (so the dimension goes from $2N \to N$), which is isomorphic to the space of operators which flip the spins, and as such, $\mc F$ plays both roles. Then $\lambda$ is replaced with the binary dot product. Thus we can characterize classical models as those for which $\lambda$, i.e. the two-form for the field space, is the binary dot product, and quantum models as those for which $\lambda$ is symplectic ( $\lambda(f,f)=0$ for all $f\in \mc F$).

\subsection{Generalized Wegner Duality}

To see the importance of the linear gauge structure, we discuss the primary use of the BrLE rules in the context of thermal properties of stabilizer systems. In Appendix \ref{ap:gw}, we give a proof of the {\it generalized Wegner duality} as first discovered by Kramers and Wannier \cite{Kramers1941} for 2D classical Ising-like models and generalized to higher dimensions by Wegner \cite{Wegner1971}.

Let $(\mc A, \omega)$ be a binary vector space with a non-degenerate two-form. Then for any subspace $\mc L\subseteq \mc A$, define the thermal partition function for $\mc L$ as 
\begin{align}
Z_{\mc L} \left(\beta; E\right)= \sum_{A\in \mc L} \exp\left(-\beta E(A)\right),
\end{align}
where $E(A)= \sum_i E_i \omega(A, \hat e_i)$ is a linear energy functional on members of $\mc A$ (linear in the real-number field, not $\mb F_2$) and $\{\hat e_i\}$ are the binary unit vectors.\footnote{If the energy functional is omitted in the arguments, it is assumed that $E_i=1$ for all $i$.} When $E_i=1$ for all $i$, we write $E(A)= \|A\|$, which represents the Hamming weight of the binary vector. The imaginary-time Generalized Wegner Duality is given by:

\begin{theorem} \label{thm:genWeg}
For any perpendicular-complement space $\mc L^{\perp_\omega}\subseteq \mc A$, the thermal partition function for inverse temperature $\beta$ and energy functional $E$ satisfies,
\begin{align}\label{eq:Wegner}
\frac{Z_{\mc L^{\perp_\omega}} \left(\beta; E\right)}{Z_{\mc A} \left(\beta; E\right)} \propto Z_{\mc L} \left(\beta'; E'\right), 
\end{align}
where $\beta' = -\ln\tanh(\beta/2)\sim 1/\beta$ and $E'$ is some linear energy functional (see Appendix \ref{ap:gw} for exact form of $E'$). If $E_i=1$, then $E'_i=1$ for all $i$.
 \end{theorem}
For the model represented by $\mc L^{\perp_\omega}$, we refer to the model represented by $\mc L$ as the {\it GW dual}.
Since the BrLE rules state $ \im \psi = (\ker \phi)^{\perp_\omega}$, the thermal model representing our stabilizer code, $\im \psi$ is GW dual to a model for $\ker \phi$, i.e. a model of its constraints. This is the same relation found in Ref. \cite{Weinstein2019} by direct computation of the trace. For $\text{TC}_2$, this implies the low-temperature model is GW dual to a high temperature model of two independent spins with infinite energy in the thermodynamic limit, since the binary vector of the constraints contains an extensive number of down-spin (1 bit) entries. In this case, the infinite energy freezes out these constraint spins at any finite temperature and we have $Z_{\im \psi_{TC_2}} \to Z_{\{0,1\}^{2L^2}}$ i.e. the toric code partition function converges to the trivial partition function of independent spins and contains no thermal phase transition. 

This $\text{TC}_2$ result is general whenever all constraints are topological i.e. the model is trivialized by changing the boundary conditions. Topological constraints are always extensive or sub-extensive and as such represent GW dual spins with infinite energy which are then frozen out at any finite temperature. Examples for which this is true include the $d=1$ Ising model and the fractonic models, $d=2$ Ising plaquette, $d=3$ Haah's cubic code \cite{Haah2011, HaahThesis} and the Cluster-cube model  (see Ref. \cite{Schmitz2019c} for detailed discussion of constraints for the last two models).  This implies that a necessary condition for such models to contain a phase transition is the existence of local constraints as demonstrated in the $d=2,3$ Ising models, the $d=3$ $\mb Z_2$ Ising gauge theory model, (so by extension the $d=3$ toric code) and the $d=3$ Ising plaquette model \cite{Espriu1997, Mueller2017}. However, the $d=3$ fractonic model, X-cube model \cite{Vijay2016}, contains local constraint but no thermal phase transition \cite{Weinstein2018}. This is a counterexample which proves local constraints alone are not sufficient for a thermal phase transition.

\subsection{Real-time Wegner Duality and Dynamical Phase Transitions}

In real-time dynamics, we define the real-time partition function as $Z^R(t)= \tr(U(t))= Z(i\beta)$. We can Wick rotate Eq. \eqref{eq:Wegner} to real time, but dividing by the trivial partition function is problematic as it can be zero. Instead, we write the general Wegner duality in real time as (for the $E_i=1$ energy functional),
\begin{align}\label{eq:Wegner2}
 Z^R_{\mc L^{\perp_\omega}} (t) \propto \sum_{A \in \mc L} &\cos^{\|\overline{A}\|}\left(\frac{t}{2}\right) \nonumber \\
 & \times (-i\sin)^{\|A\|}\left(\frac{t}{2}\right), 
\end{align}
where the proportionality includes an unimportant phase factor. We also use $\overline A$ to represent the negation of the binary vector such that $0 \leftrightarrow 1$. 

\begin{figure*}[t]
\centering
\begin{tabular}{cc}
\subfloat[{}\label{fig:DPT1}]{\includegraphics[scale=.75]{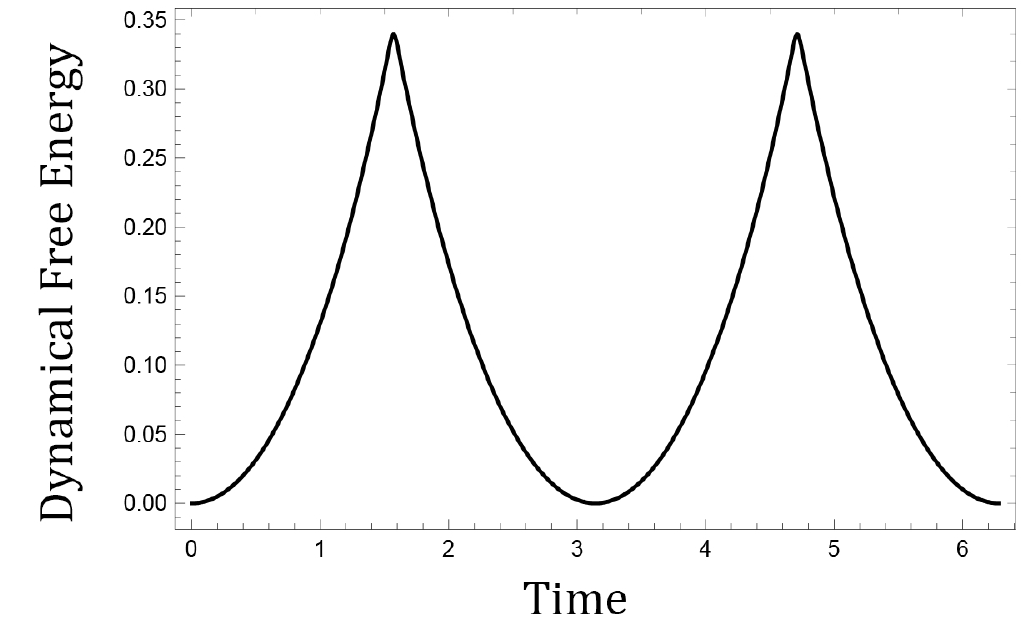}}&
\subfloat[{}\label{fig:DPT2}]{\includegraphics[scale=.75]{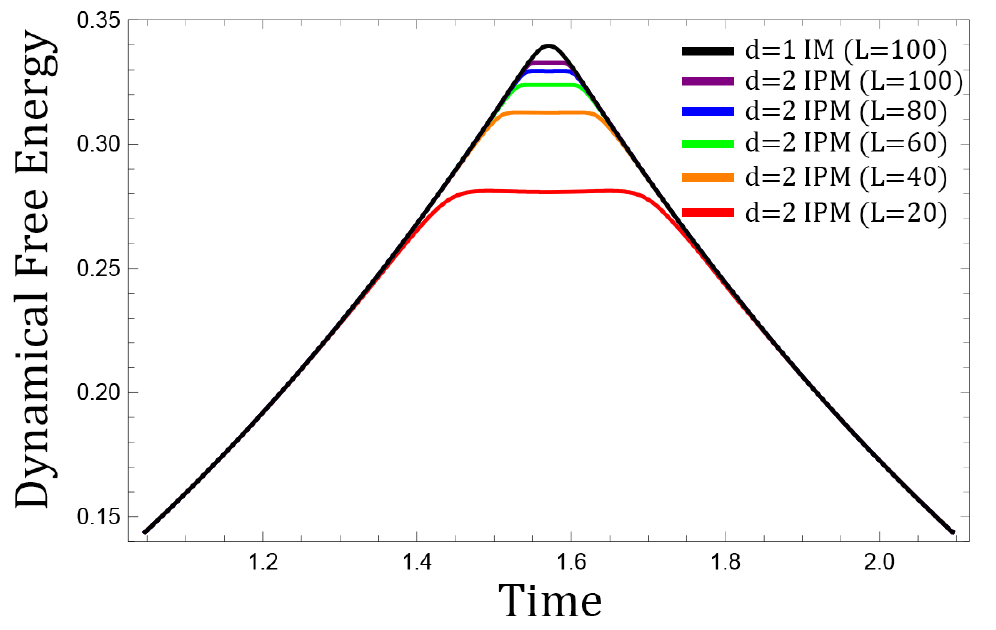}}
\end{tabular}
\caption{Plots of the dynamical free energy density at finite lattice sizes to demonstrate the formation of DPTs. (a) Plot for the $d=1$ Ising model at $L=100$. (b) A zoomed in plot about $t=\frac{\pi}{2}$ for the $d=2$ Ising plaquette model at various values of $L$. The $d=1$ Ising model is also shown to demonstrate the convergence of the Ising plaquette model DPT to that of the Ising model in the thermodynamic limit.}\label{fig:DPT}
\end{figure*}

Unlike thermal phase transitions, extensive and sub-extensive constraints do not imply the lack of a DPT, but rather are the source of many DPTs. For example, the simplest case, the $d=1$ Ising model, has the real-time partition function \cite{Heyl2015},
\begin{align}
 Z^R_{\text{Ising}_1}(t) \propto\cos^L\left(\frac{t}{2}\right) + (-i\sin)^{L}\left(\frac{t}{2}\right),
\end{align}
where the cosine term comes from the trivial constraint and the sine term from the constraint among all Ising terms. This partition function generates a DPT in the free energy density at $t_n= \frac{(2n+1)\pi}{2}$, for any integer $n$, which is the value which minimizes the modulus of the partition function. A plot of the dynamical free energy density is shown in Fig. \ref{fig:DPT1}. Importantly, this partition function only obtains a zero in the thermodynamic limit and does so at a rate of $\alpha^{L}$ for some constant $\alpha= \frac{1}{\sqrt{2}}<1$. Thus the dynamical free energy density is finite at the DPT.

We now argue that the $d=1$ Ising model (I1) DPT is generic for many models. In particular, any model which contains the extensive constraint--set of all stabilizers as denoted by $\bf{1}$ and no local constraints have a I1-type DPT. This includes all models discussed in the last section which lack a thermal phase transition. To show this, let our model contain the extensive constraint and have no local constraints. This implies the partition function can be split into
\begin{align}
Z^R \propto& Z^R_{\text{Ising}_1}(t) \nonumber \\
&+\sum_{A\neq 0, \bf{1}} \cos^{ \|\overline {A}\|}\left(\frac{t}{2}\right)(-i\sin)^{\|A\|}\left(\frac{t}{2}\right), 
\end{align}
In general, any term $A\neq 0, \bf{1} \in \ker \phi$ is strictly less than either the $0$ or $\bf{1}\in \ker \phi$ terms except at $t= \frac{(2n+1)\pi}{2}$, for any integer $n$, at which point they are equal in magnitude. Moreover by our hypothesis that there are no local constraints, $\|\overline{A}\| \sim \mc O(L)$ as well as $\|A\| \sim \mc O(L)$ for all $A\neq 0, \bf{1} \in \ker \phi$. This implies that in the thermodynamic limit, all such terms are exponentially suppressed relative to $Z^R_{\text{Ising}_1}$ and we have our result. 

This result is not surprising for a model such as $d=2$ toric code due to its connection the $d=1$ Ising model \cite{Srivastav2019}, but it is disappointing for fractonic models. One might expect that because fracton constraints have non-trivial intersection by definition \cite{Schmitz2019a, Nussinov2009a}, they might host a variety of different DPTs than that of simpler topological models. And even though this intersection does change the partition function at finite lattice sizes, all such variations are washed-out in the thermodynamic limit. 

To demonstrate, we consider the simplest fractonic model, the $d=2$ Ising plaquette model \cite{Yan2019}. This is a classical model with one Ising spin per vertex of the square lattice and a stabilizer set,
\begin{align}
S_{IPM_2}=\{b_p: p \text{ plaquettes } \},
\end{align}
where $b_p= \sigma_{p_1} \sigma_{p_2} \sigma_{p_3} \sigma_{p_4}$. We can use the $d=1$ subsystem constraints (product of stabilizers along a line) as an overly determined basis for the constraint space. Consider having one of these constraints along the x-direction and one along the y-direction. The Hamming weight of this vector is nearly $2L$ except we are over counting at the intersection by $2$. No matter where these constraints are, this intersection is always at a single plaquette, thus the actual Hamming weight is $2L-2$. So in general, if we have $a$ constraints in the x-direction and $b$ in the y-direction, the Hamming weight of the resulting vector, regardless of their exact position, is
\begin{align}
\|A(a,b)\|=& aL + bL- 2ab \nonumber \\
=&(L-a)L + (L-b)L \nonumber \\
& - 2(L-a)(L-b) \nonumber \\
=& \hlf L^2 -\hlf(L-2a)(L-2b).
\end{align}
The second equality shows that the Hamming weight of such a constraint has exactly the same value if we take $a\to (L-a), b \to (L-b)$, which corresponds to $\|A(a,b) + \bf{1}\|$. This  represents the redundancy due to dependency between the $d=1$ constraints. Thus we can sum over all $a, b \in [0, L]$ as if they are independent, and then divide by two to account for the dependency. There is also an entropic factor of $\binom{L}{a} \binom{L}{b}$ due the number of ways we can rearange these constraints. Thus we find that\footnote{Note that in spite of the tangent function being undefined at $t=n\pi$, the cosine factor out front always makes the partition function finite.},
\begin{widetext}
\begin{align}
Z^R_{\text{IPM}_2}\propto& \left(\cos\left(\frac{t}{2}\right)(-i\sin)\left(\frac{t}{2}\right)\right)^{\frac{L^2}{2}} \sum_{a,b\in [0,L]}\binom{L}{a} \binom{L}{b} (-i \tan)^{\hlf(L-2a)(L-2b)}\left(\frac{t}{2}\right),
\end{align}
\end{widetext}
 Fig. \ref{fig:DPT2} shows a plot of the dynamical free energy density for this partition function for various lattice sizes. One can see that in spite of the structure due to the intersecting constraints, the free energy density approaches that of the $d=1$ Ising model as $L$ increases. We do note that the second derivative at the value of the DPT has a different sign than that of the $d=1$ Ising model. However, it is unclear if this kind of feature can be used to detect fractonic behavior in a well-defined manner for the thermodynamic limit.    

In the cases where there are local constraints, we still have that every term $A\neq 0, \bf{1} \in \ker \phi$ is less than either the $0$ or $\bf{1}$ terms. However, these terms are not suppressed to zero in the thermodynamic limit, thus we expect a richer set of behavior for such models. We look to explore this more in future work.

\section{Quantum Quench between Stabilizer Codes}

\subsection{Derived Linear Gauge Structures}

In this section, we discuss some new linear gauge structures which are derived from one or two other linear gauge structures. These are useful for discussing quantum quenches between different stabilizer codes. 

\begin{figure}[t]

\centering

\includegraphics[scale=.35]{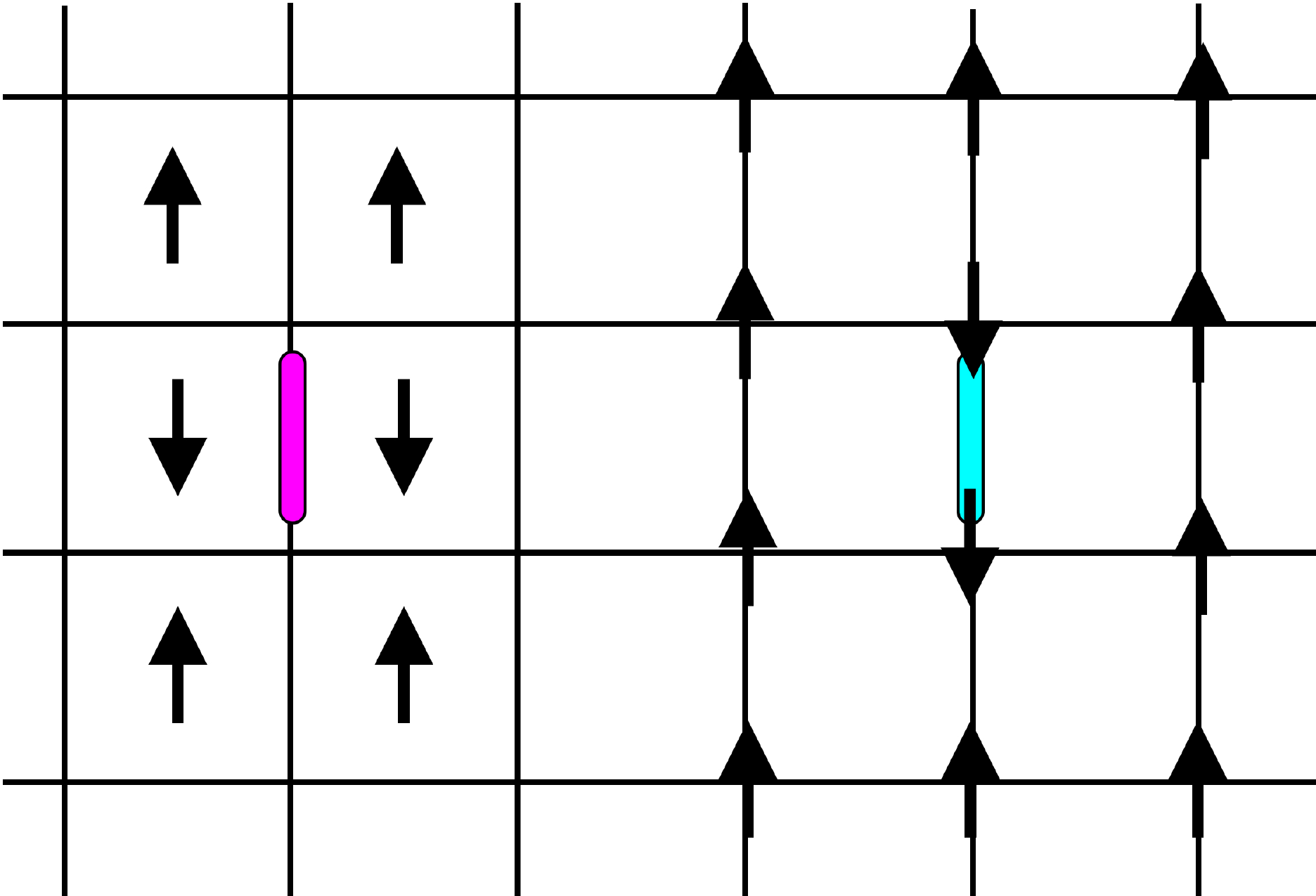}

\caption{depiction of how single-qubit X and Z operators in the $d=2$ toric code are mapped onto $d=2$ Ising model terms. Magenta represents X-type operator support and cyan represents Z-type operator support.}\label{fig:TCtrans}

\end{figure}

Our first case of a classical model derived from a stabilizer code is the {\it transpose} gauge structure. The transpose simply reverses the roles played by the two vector spaces, i.e. $\text{GS}^\top=\left((\mc F, \lambda, \psi), (\mc A, \omega, \phi)\right)$, which also satisfies the braiding relation by Eq. \eqref{eq:stabgauge}.\footnote{One simply uses the symmetry of the these forms to flip the positions of the entries of $\omega$ and $\lambda$.} To understand how this represents a classical spin model, note the interpretation of a gauge structure is that the potential space represents virtual degrees of freedom, and the field space represents physical degrees of freedom. Applying this to the transpose, each stabilizer is attach to a physical classical spin, i.e. for $\text{TC}_2$ our virtual vertex and plaquette spins now represent the physical degrees of freedom of the system. To construct terms in our Hamiltonian, we do the same as we did with the stabilizer code where virtual and physical degrees of freedom are related by $\phi$. For the transpose, this connection is analogously provided by $\psi$. Every member of our basis for $\mc F$, namely single qubit X and Z Pauli operators, now represent terms in the Hamiltonian as is given by
\begin{align}\label{eq:transHam}
H_{\text{GS}^\top}=\hlf \sum_i\left(\left(1- \psi(x_i)\right)+\left(1-\psi(z_i)\right)\right),
\end{align}
This again can be reduced to a classical energy functional for members of $g\in \im \phi$, which by analog, represents the excitations of this model.
For $\text{TC}_2$, we know that single-qubit operators generate pairs of excitation, in which case for edge $e$, $\psi(x_e)\simeq \sigma_{p_{e_1}}\sigma_{p_{e_2}}$ and $\psi(z_e)\simeq \sigma_{v_{e_1}}\sigma_{v_{e_2}}$, where $\sigma$ represents the physical spins of the transpose. This is demonstrated in Fig. \ref{fig:TCtrans}. Thus the transpose of $\text{TC}_2$ is two independent copies of the $d=2$ Ising model \cite{Dennis2002,Cobanera2011, Nussinov2008}. That the X-generated terms and Z-generated terms of Eq. \eqref{eq:transHam} are independent is a general feature for CSS stabilizer codes. We refer to the factors of a transpose CSS model as the {\it X-sector} and {\it Z-sector}, respectively. 

The transpose model has two important connections to its stabilizer code partner. In the condensed matter context, it
represents the classical model which can be gauged to form the gauge theory representing the stabilizer code \cite{Vijay2016, Williamson2016, Kubica2018, Nussinov2008, Cobanera2011}. In the context of QECC, the transpose represents an important model for characterizing the protection of information in the ground space. In Ref. \cite{Dennis2002}, it is argued that a thermal phase transition in the {\it random-bond} version of the transpose model represents the theoretical upper-bound on the decoding rate, i.e. the critical temperature represents the robustness of information in the ground space. This connection is reviewed in Appendix \ref{ap:trans}. The partition function of the random-bond transpose model is given by
\begin{align}\label{eq:transposeZ}
Z_{\text{GS}^\top}(\beta)=&\frac{1}{Z_{\mc F}(\beta)} \sum_{f \in \mc F}\exp\left(-\beta\|f\|\right) \nonumber \\
& \times  \sum_{g \in \im \phi} \exp\left(-\beta \|f+g\|)\right).
\end{align}
The second sum in this definition comes from the fact that the projectors of the random-bond version of Eq. \eqref{eq:transHam} are allowed to be for $\pm1$ eigenstates with a Gibbs-like probability distribution (see Appendix \ref{ap:trans} for details).

As the transpose is a gauge structure, we can apply the GW duality to it as well. The relevant BrLE rule is $\im\phi= (\ker \psi)^{\perp_\lambda}$, so the GW dual of the standard transpose model (not random-bond) is a model for $\ker \psi \supseteq \im \phi$. Furthermore as $\ker \psi/\im \phi$ represents logical operators, the minimum support of operators in this set defines the code distance\cite{Terhal2015}. If the code distance scales as $\mc O(L)$, i.e. it is a ``good'' code, contributions coming from $\ker \psi/\im \phi$ represent infinite energy dual spins and so in the thermodynamic limit, the GW duality reads,
\begin{align}
\frac{ Z_{\im \phi}(\beta)}{ Z_{\mc F}(\beta)} \to Z_{\im \phi}(\beta'). 
\end{align}
That is, the transpose model is generally GW dual to itself. This is a reflection of the fact that every stabilizer represents a constraint of the transpose model. For our example, the constraints of the $d=2$ Ising model are all closed loops of edges, exactly the operators of the $d=2$ toric code. 

Though we have only showed the GW duality for the standard version of the transpose, we show in Appendix \ref{ap:trans} using similar methods that
 \begin{align}\label{eq:Tdual}
 \frac{Z_{\text{GS}^\top}(\beta)}{\mc Z_{\mc F}(\beta)} \propto Z_{\im \phi}(2 \beta').
 \end{align}
So study of the random-bond version of the model can be reduced to study of the standard version. That any transpose of a locally generated stabilizer code is self-dual suggests that the transpose generically contains a thermal phase transition and gives insight into the robustness of information in the code space. For CSS codes, this self-duality is expressed as the X-sector being dual to the Z-sector and vis versa. 

The second derived gauge structure is the {\it composite}. Suppose we have the two gauge structures,  GS$=\left((\mc A, \omega, \phi),(\mc F, \lambda, \psi)\right)$ and $\text{GS}'=\left((\mc A', \omega', \phi'),(\mc F, \lambda, \psi')\right)$ which have the same field space (live on the same qubits). We then form the composite as, $\text{GS} \circ \text{GS}'=\left((\mc A, \omega, \psi'\phi),(\mc A', \omega', \psi\phi')\right)$, where for all $A\in \mc A$ and $A' \in \mc A'$,
\begin{align}
 \omega'(\psi'\phi(A), A')=& \lambda(\phi(A), \phi'(A')) \nonumber \\
 =& \omega(A, \psi \phi'(A')),
\end{align}
which we recognize as the braiding condition for the composite. The BrLE rules are then
 \begin{subequations}\label{eq:conss}
\begin{align}
(\ker \psi' \phi)^{\perp_\omega} =& \im \psi \phi', \label{eq:brle1}\\
(\im \psi'\phi)^{\perp_{\omega'}} =& \ker \psi \phi' \label{eq:brle2},
\end{align}
\end{subequations}
For stabilizer codes, composites represent classical spin models. $\im \psi_{TC_2} \phi_{X}$ simply represents the excitations of the Z-type stabilizers. More interestingly, consider $\im \psi_{X} \phi_{TC_2}$. X-type $a_v$ operators map into $\im \phi_X =\ker \psi_X$ so they all map to $0$ under $\psi_{X} \phi_{TC_2}$. On the other hand, the Z-type $b_p$ operators anti-commute with the X-type operator on every edge in its support. Thus $\psi_{X} \phi_{TC_2}$ kills the vertex operators and maps the plaquettes to their boundary and as such, $\im \psi_{X} \phi_{TC_2}$ represents the same space as the X-sector of the transpose gauge structure. These properties are general for the composites between a CSS stabilizer code and either the trivial X or Z stabilizer codes. That is, { \bf $ \im \psi_X \phi$ and $\im \psi_Z \phi$ represents the X- and Z- sectors of the transpose where $\phi$ representing a CSS code.}

Another way to combine two stabilizer codes with the same field space is by forming the {\it subsystem} gauge structure,\footnote{The name is derived from the fact that the resulting gauge structure has the from of a subsystem code \cite{Terhal2015}.} $\text{GS}\&\text{GS}' =\left(\mc A \oplus \mc A', \omega'', \Phi), (\mc F, \lambda, \Psi)\right)$. $\omega''= \omega \oplus \omega'$ i.e. it is the natural two-form on $\mc A \oplus \mc A'$ and $\Phi(A \oplus A')= \phi(A) + \phi'(A')$ for all $A\oplus A' \in \mc A \oplus \mc A'$. $\Psi$ can be derived via the braiding condition,
\begin{align}
\lambda(\Phi(A \oplus A'), f)=& \lambda(\phi(A), f) + \lambda(\phi'(A'),f) \nonumber \\
=& \omega(A, \psi(f)) + \omega'(A', \psi'(f)) \nonumber \\
=& \omega''(A \oplus A', \psi(f) \oplus \psi'(f)).
\end{align}
Therefore, $\Psi(f) =\psi(f) \oplus \psi'(f)$, for all $f \in \mc F$. This immediately implies the additional BrLE rule
\begin{align}\label{eq:SSBrle}
(\ker \Phi)^{\perp_{\omega''}} = \im \Psi,
\end{align}
where we note that $\ker \Phi$ represents all stabilizer products shared between the two codes. That is if $A\in \mc A$ and $A' \in \mc A'$ are such that $ \phi(A)= \phi'(A')$, then $A\oplus A' \in \ker \Phi$.

The subsystem and composite gauge structure have a useful connection to each other. Consider the set of all $J \oplus J' \in \im \Psi$ such that $J$ is constant. This is generated by all $f \in \mc F$ such that $\psi(f) =J$, and any two such $f_1, f_2$ are such that $\psi(f_1+f_2)=0$. Thus given such an $f$, our set is $J \oplus \psi'[f +\ker \psi]= \Psi(f)+ 0\oplus\psi'[\ker\psi] \supseteq \Psi(f) + 0\oplus\im \psi' \phi$. Thus our set is an affine subspace, i.e. is a subspace shifted by the constant $J$ and is isomorphic to a superset of $\im \psi'\phi$.

\subsection{Quantum Quenches at Zero Temperature}

Practically speaking, one can measure the real-time partition function by measuring the Loschmidt echo, $\braket{\psi|\exp(-i Ht)|\psi}$, of a quantum quench. However, the starting state $\ket{\psi}$ must be an even superposition in magnitude over all energy eigenstates. In practice, such a state can be a challenge to manage exactly. However, a quantum quench between two different stabilizer codes can achieve this for a subspace corresponding to composite and subsystem gauge structures. To show this, consider the overlap between stabilizer states of $S$ and $S'$, defined by\footnote{The overlap is roughly like ``$|\braket{J|J'}|^2$'' but taking the trace over the projection operators avoids an arbitrary choice of logical state.}
\begin{align} \label{eq:2trace}
 \tr&\left(p^{(J)}_S p^{(J')}_{S'}\right) \nonumber \\
\propto& \sum_{A\in \mc A} \sum_{A' \mc A'} (-1)^{\omega(A,J)} (-1)^{\omega'(A' J')} \tr\left(\phi(A) \phi'(A')\right)\nonumber \\
=& \sum_{A\oplus A' \in \ker \Phi} (-1)^{\omega''(A \oplus A', J \oplus J')} \nonumber\\
\propto& [ J\oplus J' \in (\ker \Phi)^{\perp_{\omega''}}], 
\end{align}
where we have used the fact that all Pauli operators--except the identity-- are traceless and Eq. \eqref{irrepID} from Appendix \ref{ap:gw} and where $[]$ is the Iverson bracket (evaluates to $1$  if the statements is true and $0$ otherwise). We then apply the BrLE rule in Eq. \eqref{eq:SSBrle} to show that this overlap is non-zero if and only if $J \oplus J' \in \im\Psi$ and a constant for all such $J \oplus J'$. So a stabilizer state of $S'$ as labeled by $J'$ is an even superposition of stabilizer states of $S$ as labeled by members of $\im \Psi$ for which $J'= \psi'(f)$ is a constant. As discussed in Section \ref{sec:LGS}, this set is  $\Psi(f)+ \psi[\ker\psi']\oplus 0$. $p^{(J')}_{S'}$ represents a pure-state projection operator if and only if $\ker \psi'= \im \phi'$ and thus a zero temperature quench ($J'=0$) yields
\begin{align}\label{eq:0quench}
|\braket{\psi'(0)| \exp\left(-i H_S t\right)|\psi'(0)}|^2 \propto \left| Z^R_{\im\psi \phi'}(t)\right|^2
\end{align}

 One can then apply Eq.\eqref{eq:Wegner2} to make this a sum over $\ker\psi'\phi$ using the BrLE rules for the composite gauge structure. So for $S'= S_X$ and $S= S_{TC_2}$, our quantum quench corresponds to the real-time partition function for $\im \psi_{TC_2}\phi_X$, which as discussed before represents all $b_p$ excitations and has a I1 DPT. If we flip the roles of the stabilizers, i.e. start with a toric code grounds state and evolving using the trivial traverse-field Hamiltonian, we have the real-time partition function for $\im\psi_X \phi_{TC_2}$ which corresponds to  the $d=2$ Ising model, which has a different DPT \cite{Heyl2013}. This suggests a kind of cross-over between these two quenches similar to that studied in Ref. \cite{Srivastav2019}.

\subsection{Thermal Stability of the Quantum Quench for a CSS Code} \label{sec:Tstable}

 As  with the example, an important quench is when $S' =S_X$ or $S_Z$ and $S$ is a CSS stabilizer code which we refer to as a {\it CSS quench}. We now arrive at our primary result, a zero-temperature DPT in a CSS quench is robust to thermal noise, if the information protection of the underlying code is robust, i.e. the transpose contains a phase transition. For simplicity, let $S' =S_X$, but all analogous statements hold for $S_Z$. We evaluate thermal stability by considering the {\it thermal-quantum quench auto-correlator} (TQQAC) defined as
 \begin{widetext}
 \begin{align}
 Y^2_{(S, S')}(t; \beta) =& \tr \left(  \rho(0; \beta)\rho(t; \beta) \right)\nonumber \\
 = & \frac{1}{Z^2_{\im \psi'}(\beta)}\tr\left(\exp\left(-\beta H_{S'}\right) \exp\left(-it H_S\right) \exp\left(-\beta H_{S'}\right) \exp\left(itH_S\right)\right) \nonumber \\
 =&\frac{1}{Z^2_{\im \psi'}(\beta)} \sum_{J_1', J_2' \in \im \psi'} \exp\left(-\beta(\|J_1'\| +\|J_2'\|) \right) G^2_{(S, S')}(t; J'_1, J'_2),
 \end{align}
 \end{widetext}
where $G^2_{(S, S')}(t; J'_1, J'_2)$ are the Green functions as defined in Appendix \ref{ap:quench} for all $J_1', J_2' \in \im \psi'$. Though study of the TQQAC differs from the generalizations to the Loschmidt echo discussed in Refs. \cite{Heyl2017, Heyl2017a,Bhattacharya2017, Lang2018, Abeling2016}, it does possess the correct asymptotic behavior one would expect. That is, at zero temperature, it reduces to the  modulus-squared of the pure-state Loschmidt echo and at infinite temperature, it reduces to a constant. The key for a CSS quench is that we can simplify the Green functions. In general, the Green function is zero unless there exists an $A\in \mc A$ such that $J_1'+J_2'= \psi'\phi(A)$. For $S= S_{TC_2}$ and $S'= S_X$, this implies that if we view $J_1^X, J_2^X$ as two collections of open strings, then the associated Green function is non-zero only if they have the same endpoints as shown in Fig. \ref{fig:quenchloop}. In general for a CSS quench ($S'= S_X$), we show in Appendix \ref{ap:quench} that when this condition is satisfied,
\begin{widetext}
\begin{align} \label{eq:CSSgreen}
G^2_{(S, S_X)}(t; J_1^X, J_2^X) \propto \left| \sum_{J\in \im \psi\phi_X} (-1)^{\omega(A, J)} \exp\left( -i t\|J\|\right) \right|^2 =  \left|Z^R_{\im \psi\phi_X}(t; A)\right|^2.
\end{align}
\end{widetext}
When we apply the real-time GW duality to $Z^R_{\im \psi \phi_X}(t;A)$, the extra sign in the sum only has the effect of shifting the sum in Eq. \eqref{eq:Wegner2} to an affine subspace so that
\begin{widetext}
\begin{align} \label{eq:ZofA}
Z^R_{\im \psi\phi_X}(t; A) \propto \sum_{C \in \ker \psi_X\phi} \cos^{\|\overline{A+C}\|}\left(\frac{t}{2}\right)(-i\sin)^{\|A+C\|}\left(\frac{t}{2}\right).
\end{align}
\end{widetext}
Note that here we have a similar situation as the discussion around DPT's. In cases where $0, \bf{1}$$ \in \ker \psi \phi_X$, i.e. for a I1 DPT, $Z^R_{\im \psi\phi_X}(t; A)$ is largest for all $t \neq t_n$ when $A \in \ker\psi_X\phi$. Because of this, we can characterize $Z^R_{\im \psi\phi_X}(t; A) $ as a local quantity in the following way: for $A=0$ (equivalent to all spins up in the transpose model) $Z^R_{\im \psi\phi_X}(t; 0)= Z^R_{\im \psi\phi_X}(t)$, i.e. it takes on the zero-temperature value, and $|Z^R_{\im \psi\phi_X}(t; A)|^2 \sim \alpha(t)^{\|A\|}$ for some $\alpha(t) < 1$, i.e. it's value falls of exponentially with the size of $A$.

Using this form of the Green functions, we can shift the sum to find
\begin{widetext}
\begin{align}
Y^2_{(S, S_X)}(t, \beta) \propto& \frac{1}{Z^2_{\mc A_X}(\beta)} \sum_{A_X \in \mc A_X} \exp\left(-\beta \|A_X\|\right) \sum_{A \in \mc A} \exp\left( -\beta \|A_X + \psi_X\phi(A)\|\right) \left| Z^R_{\im\psi\phi_X}(t; A)\right|^2 .
\end{align}
\end{widetext}
So the TQQAC is a weighed sum over Green functions, with a Boltzmann weight as given by the set $ A_X +\im \psi_X\phi$.  As $\im \psi_X\phi$ always represents the X-sector of the transpose of our CSS code, we have the relation
\begin{align}
\frac{Z_{\mc A_X}(\beta)}{Z_{\text{GS}^\top_X}(\beta)} Y^2_{(S, S')}(t, \beta) \propto \braket{ \left| Z^R_{\im\psi\phi'}(t)\right|^2}_{\text{GS}^\top_X}.
\end{align}
That is, the normalized TQQAC is the expectation value of the Green functions in the X-sector random-bond version of the transpose model of our CSS code. In general, we expect the random-bond transpose thermal model in either the X or Z sector to have an ordered phase at temperatures below the critical temperature as determined by the theoretical decoding rate of the stabilizer code. Therefore, any DPT in the zero-temperature Green function is robust below this critical temperature, if such a critical temperature exists. Moreover, we can see that for $t= t_n$, i.e. at the time of a DPT, all Green functions become equal and  $Y^2_{(S, S')}(t_n, \beta) \propto Z_{\text{GS}^\top_X}(\beta)/Z_{\mc A_X}(\beta)$. So we can use the height of the DPT as a function of temperature for a CSS quench to extract the transpose model free energy density.
\begin{figure}

\centering

\includegraphics[scale=.15]{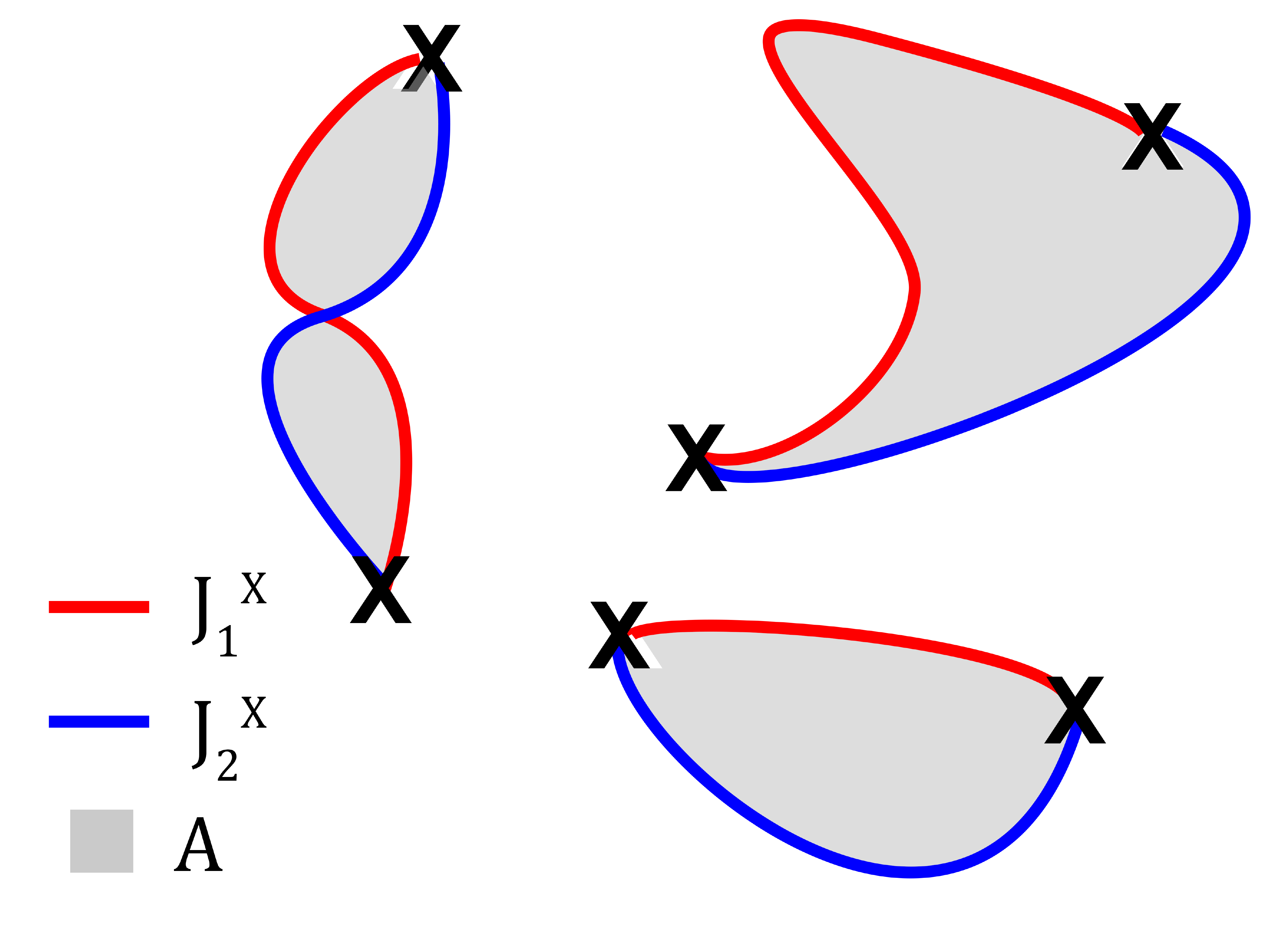}

\caption{Example for $\text{TC}_2$ of a non-zero Green function which satisfies $J_1^X+J_2^X= \psi_X\phi_{TC_2}(A)$.}\label{fig:quenchloop}

\end{figure}

Clearly thermal stability holds for a CSS quench of our $d=2$ toric code example. And even though the $d=3$ fractonic models, X-cube, Haah's cubic code, and cluster-cube only contain I1-type DPTs in either sector, these should all be robust to thermal noise for a CSS quench. For example, the X-sector transpose of the X-cube model (cube excitations; see Refs \cite{Vijay2016, Prem2017c}) is the $d=3$ Ising plaquette model which as previously discussed has a thermal phase transition. 

Note if we exchange the roles of the two stabilizer codes, we should generally expect any DPT to be lost when the X (Z)-sector of the stabilizer code for $S$ does not contain a thermal phase transition. This is the case for $d=2$ toric code, but is not the case for some examples, namely the X-sector of the $d=3$ toric code which is equivalent to $d=3$ Ising gauge theory.  

\section{Conclusions} 

In this paper, we have discussed the general theory of DPTs for stabilizer codes, and established a connection between the properties of the QECC and the thermal stability of DPTs. In particular, we show a thermal-quantum quench  between a product thermal state and a CSS stabilizer code demonstrates a robust DPT up to the critical temperature corresponding to the resilience of information in the code space in dimensions $d>1$. We also show a connection between DPTs and a thermal phase transition in these models by demonstrating that generally, stabilizer models without a thermal phase transition have a DPT which is indistinguishable from that of the $d=1$ Ising model. This includes many three-dimensional fracton models.  

This work opens up many possible avenues of further study. The simplest is a characterization of DPTs for models with local constraints i.e. those which are distinct from the $d=1$ Ising model. This could lead to a better understanding of the  connection between DPTs and thermal phase transitions. Another possibility is to consider more exotic quenches between different stabilizer codes in search of different classes of DPTs. For example, the results of Ref. \cite{Schmitz2019c} could be used to devise quenches which result in a partition function of fractonic loops in three-dimensions. The natural question then arises if the fractonic nature of these loops results in DPTs which are qualitatively distinct from that of the $d=3$ Ising gauge theory which contains freely deformable loops.  We also look to use a similar formalism to explore the robustness of DPTs when perturbed beyond the stabilizer code limit. 

\section{Acknowledgments}

The author would like to especially thank Rahul M. Nandkishore for advice on both the content and structure of the manuscript, as well as Victor Gurarie and Leo Radzihovshy for their comments on the methods of this paper. This material is based upon work supported by the Air Force Office of Scientific Research under award number FA9550-17-1-0183. 

\appendix

\section{Proof of the Generalized Wegner Duality}\label{ap:gw}

In this Appendix, we look to prove Theorem \ref{thm:genWeg}. These results generalize those of Ref. \cite{Wegner1971} by abstracting away any notion of a lattice or spin model to that of the space $(\mc A, \omega)$, and generalizing the non-degenerate two-form $\omega$ beyond just the binary dot-product. To be clear, non-degeneracy is the property that for any $A \in \mc A$, if $\omega(A,B)=0$ for all $B\in \mc A$, then $A=0$, which can also be stated as $\mc A^{\perp_\omega}=\{0\}$.

For the space $(\mc A, \omega)$, we draw on results from Ref. \cite{Bernhard2009} which finds that any non-degenerate two-form is isomorphically equivalent to any other. That is, for any two non-degenerate forms $\omega$ and $\omega'$ on the same space $\mc A$, there exists an automorphism, $\alpha$ such that $\omega(A,B) = \omega'(\alpha(A), B)$. For our purposes, this equivalence can be expressed in terms of the irreducible representations (irreps) of the addition group of $\mc A$. The function $\omega(A,*) :\mc A \to \mb F_2$ represents an irrep of the addition group as indexed by $A$. So $\omega(\alpha(A), *)$ is just a re-indexing of the irreps. This allows us to use the following identity for any $(\mc A, \omega)$ and any subspace $\mc L \subseteq \mc A$ \cite{Tinkham2003},
\begin{align}\label{irrepID}
\frac{1}{|\mc L|} \sum_{A \in \mc L} (-1)^{\omega(A, B)}= [B \in \mc L^{\perp_\omega}],
\end{align}
As all of $\mc L^{\perp_\omega}$ indexes the trivial representation of addition in the subgroup $\mc L$. 

We now apply the identity of Eq. \eqref{irrepID} to the partition function,
\begin{align}
Z_{\mc L^{\perp_\omega}}(\beta; E) =& \sum_{B \in \mc L^{\perp_\omega}} \exp\left(-\beta E(B)\right) \nonumber \\
\propto& \sum_{A \in \mc L} \sum_{B \in \mc A} (-1)^{\omega(A, B)}\exp\left(-\beta E(B)\right).
\end{align}
We can preform the sum over $\mc A$ which factors due to the linearity of $E$ and $\omega$. Let $\alpha$ be the automorphism which takes $\omega$ to the binary dot product, so that
\begin{align}
\sum_{B \in \mc A}& (-1)^{\omega(A, B)}\exp\left(-\beta E(B)\right) \nonumber \\
&= \prod_i \left(1 + (-1)^{\alpha(A)_i} \exp\left(-\beta E_i\right)\right),
\end{align}
For each factor as indexed by $i$, we can factor out $\left(1 +\exp\left(-\frac{\beta}{2} E_i\right)\right)$, which collectively forms $Z_{\mc A}(\beta; E)$. Then define $\beta'= -\ln \tanh\left(\frac{\beta}{2}\right)$ and $E_i'= \frac{\ln \tanh\left(\frac{\beta E_i}{2}\right)}{\ln \tanh\left(\frac{\beta}{2}\right)}$. We thus recognize 
\begin{align}
Z_{\mc L^{\perp_\omega}}(\beta; E)\propto Z_{\mc A}(\beta; E) Z_{\mc L}(\beta'; E'),
\end{align}
where $E'(A)= \sum_i E_i' \alpha(A)_i= \sum_i E'_i \omega(A, \hat e_i)$ and $\hat e_i$ are the binary unit vectors. Clearly if $E_i=1$ then $E'_i=1$ for all $i$. 
 
\section{Derivation of the Transpose Partition Function}\label{ap:trans}

In this Appendix, we look to demonstrate the connection between the random-bond transpose partition function of Eq. \ref{eq:transposeZ} for the stabilizer code $S$ and the theoretical upper-bound on the decoding rate of the code, generalizing the results of Ref.\cite{Dennis2002}.

First we must consider our error model. For any quantum channel, we can expand in the Krauss operator form, and expand the Krauss operators in the Pauli operator basis. Because we are always measuring Pauli operators of our stabilizer set, all ``off-diagonal'' contributions to the channel are projected out and we are effectively left with a {\it Pauli channel} i.e. all Krauss operators are proportional to Pauli operators. We shall assume the Pauli-Krauss operators of our error channel are such that our X and Z single-qubit operators are applied iid with a probability $p \in [0,1]$.\footnote{Note this implies Y single qubit operators are applied with probability $p^2$ which may not be that realistic. For CSS codes this is still a reasonable approximation, however of non-CSS codes, one may have to modify the error model.} Then the probability that any $f \in \mc F$ is applied to our system for this error model is 
\begin{align}
\prob(f) =& (1-p)^{2N} \left(\frac{p}{1-p}\right)^{\|f\|} \nonumber \\
 \propto& \exp\left(-\beta \|f\|\right),
\end{align}
where we define 
\begin{align}\label{ap:NMline}
\exp\left(-\beta\right) = \frac{p}{1-p}.
\end{align}
Now suppose $p$ is known, and we measure our syndrome (excitation pattern) to be $J\in \im \psi$. For simplicity, suppose the measurements themselves contain no errors. Decoding this syndrome into a Pauli error is not unique, as for a given error $f$, all of  $f +\ker\psi$ generates the same syndrome. So we assume that the decoder is such that we choose $f+g \in f+ \ker\psi$ with a probability proportional to $\exp\left(-\beta \|f+g\|\right)$. If we have chosen a correct decoding, $f+g \in f+ \im \phi$, which is the case with probability, 
\begin{widetext}
\begin{align}\label{correct}
\prob(\text{ We correctly decode }J) = \left(\frac{\sum_{g \in \im \phi}\exp\left(-\beta \|f+g\|\right)}{\sum_{g' \in \ker \psi}\exp\left(-\beta \|f+g'\|\right)}\right).
\end{align}
\end{widetext}
In the thermodynamic limit i.e. the idealized limit of the code, if $f$ is of finite size, then this probability is always 1 as the only difference between the top and bottom sum is the terms coming from the logical operators, and in that case, they represent infinite energy. However, this does not tell us at what temperature $f$ is likely to be of finite size. To answer this, we consider the partition function of Eq. \ref{eq:transposeZ} which is a weighed sum of the numerator in Eq. \eqref{correct}. A local order parameter for this model is represented by the expectation value of the ``field'' or the average value of the spins representing the eigenvalue of a given stabilizer. If there is an ordered phase, the spin is on average more likely to be pointing up (be measured with a +1 eigenvalue), and as we know from renormalization arguments, downward spins tend to clump in local configurations. Thus most spins are pointing up and those which are not form local patterns. This implies we are far more likely to find that $f$ in the above expression is of finite support and we decode with asymptotically perfect precision. In the disordered phase, every spin is as likely to be up as down and all sense of locality of such configurations is lost. Thus $f$ in the above is just as likely to be finite as it is to not be finite and the probability of correctly decoding goes as roughly $\frac{1}{4^{d_\ell}}$, where $d_\ell$ is the number of logical qubits in the code. Thus, a phase transition in the transpose spin model represents an upper bound on the rate of error one can tolerate and still decode, where the relation between the probability and critical temperature is given by equation \eqref{ap:NMline}. 

To be clear, Eq. \eqref{eq:transposeZ} is a random-bond version of the transpose model in the sense that the Hamiltonian of Eq. \eqref{eq:transHam} is altered by allowing the sign of the projections to be either $\pm$ with some probability. For a fixed instance of a random-bond Hamiltonian, we can represent the energy functional of this Hamiltonian by forming a constant vector $f\in F$ with a 1 for every term with a $+$ and a 0 for every term with a $-$. Then the energy functional for this instance is parameterized by $f$ and given by $E_f(g) = \|f+g\|$. Eq. \eqref{eq:transposeZ} then weights the partition functions of each specific Hamiltonian instance with a probability factor $\propto\exp(-\beta \|f\|)$. 

We also show Eq. \eqref{eq:Tdual} by applying the identity of Eq. \eqref{irrepID} to Eq. \eqref{eq:transposeZ} and shifting the sum $g \to f+g$,
\begin{widetext}
\begin{align}
Z_{\mc F}(\beta) Z_{\text{GS}^\top}(\beta)\propto& \sum_{h \in \ker \psi} \sum_{f,g \in \mc F}(-1)^{\lambda(g,h)} \exp \left(-\beta \|f\|\right) \exp\left(-\beta \|f+g\|\right) \nonumber \\
=&\sum_{h \in \ker \psi} \left(\sum_{f\in \mc F} (-1)^{\lambda(f,h)} \exp \left(-\beta \|f\|\right)\right)^2= Z^2_{\mc F}(\beta)\sum_{h \in \ker \psi} \exp\left(-2\beta' \|h\|\right) \nonumber \\
=& Z^2_{\mc F}(\beta) Z_{\ker \psi}(2\beta'),
\end{align}
\end{widetext}
using the same methods as used in Appendix \ref{ap:gw}. Finally for good codes with a code distance that scales as $\mc O(L)$,  $Z_{\ker \psi}(2\beta') \to  Z_{\im \phi}(2\beta')$ in the thermodynamic limit. 

\section{Detailed Calculations for the TQQAC} \label{ap:quench}

In this Appendix, we discuss some of the detailed derivations surrounding the discussion in Section \ref{sec:Tstable}. We start by defining the Green functions used in the expansion of the TQQAC,
\begin{widetext}
\begin{align}\label{Gdef}
G^2_{(S, S')}(t; J'_1, J'_2) =& \tr\left(p_{S'}^{(J'_1)} \exp\left(-it H_S\right) p_{S'}^{(J'_2)}\exp\left(itH_S\right)\right) \nonumber \\
=& \sum_{J_1, J_2 \in \im \psi} \exp\left(-it \left(\|J_1\| -\| J_2\| \right)\right) \tr \left(p^{(J_1')}_S p^{(J_1)}_{S'}p^{(J_2')}_S p^{(J_2)}_{S'} \right).
\end{align}
\end{widetext}
Focusing on the trace,
\begin{widetext}
\begin{align}
\\tr \left(p^{(J_1')}_S p^{(J_1)}_{S'}p^{(J_2')}_S p^{(J_2)}_{S'} \right) \propto & \sum_{A, B\in \mc A} \sum_{A', B'\in \mc A'}(-1)^{\omega(A,J_1)}(-1)^{\omega(B, J_2)} (-1)^{\omega'(A',J'_1)}(-1)^{\omega'(B', J'_2)} \nonumber \\
& \times \tr\left( \phi(A) \phi'(A') \phi(B) \phi'(B') \right)
\end{align}
\end{widetext}
Just as with Eq. \eqref{eq:2trace}, the operator inside the trace must be the identity to be non-zero, but we also have to consider the phase generated by the commutation of operators. So in total, the following must be true for the trace to be non-zero:
\begin{subequations}
\begin{align}
\lambda(\phi(A), \phi'(A'))=& \lambda(\phi(A), \phi'(B')) \nonumber \\
= \lambda( \phi(B), \phi'(A'))&= \lambda(\phi(B), \phi'(B')), \\
\phi(A+B) = &\phi'(A'+B').
\end{align}
\end{subequations}
 As these are true for all nonzero terms, the commutation conditions ($\lambda$ conditions) imply that for all $A'\in \mc A'$ and $A,B \in \mc A$,
\begin{align}
0=&\lambda(\phi(A), \phi'(A')) + \lambda(\phi(B), \phi'(A')) \nonumber \\
=& \lambda(\phi(A+B), \phi'(A'))=\omega(A+B, \psi \phi'(A') ).
\end{align}
Therefore, $A+B \in (\im \psi \phi')^{\perp_\omega}$, and likewise $A'+B' \in (\im \psi' \phi)^{\perp_{\omega'}}$.  To simplify the sums, we shift $B \to A +B$ and $B' \to A'+B'$, and include the possible phase of the trace as given by $\omega(A, \psi\phi'(A'))$, such that
\begin{widetext}
\begin{align}
\tr \left(p^{(J_1)}_S p^{(J_1')}_{S'}p^{(J_2)}_S p^{(J_2')}_{S'} \right) \propto & \sum_{A \in \mc A} \sum_{A' \in \mc A'}(-1)^{\omega(A, J_1 +J_2 + \psi \phi'(A'))} (-1)^{\omega'(A', J'_1 + J'_2 )} \sum_{B\oplus B' \in \ker \Phi }(-1)^{\omega''(B\oplus B', J_2 \oplus J'_2)},
\end{align}
\end{widetext}
For the sums over $\mc A$ and $\mc A'$, we have
\begin{align}
\sum_{A \in \mc A}& \sum_{A' \in \mc A'}(-1)^{\omega(A, J_1 +J_2 + \psi \phi'(A'))} (-1)^{\omega'(A', J'_1 + J'_2 )} \nonumber \\
\propto & \sum_{A' \in \mc A'}[J_1 +J_2= \psi\phi'(A')] (-1)^{\omega'(A', J'_1 + J'_2 )},
\end{align}
To evaluate the final sum, let $C'\in \mc A'$ be any vector such that $ \psi \phi'(C')= J_1+ J_2$. If no such $C'$ exists, then the sum is zero. Now all other $A'\in \mc A'$ for which $ \psi \phi'(A')= J_1+ J_2$ are such that $\psi\phi'(A'+C') =0$. Thus the sum is over the affine subspace $C' +\ker\psi\phi'$. So we shift the sum by $C'$ so that 
\begin{widetext}
\begin{align}
 \sum_{A' \in \mc A'}&[J_1 +J_2= \psi\phi'(A')] (-1)^{\omega'(A',J'_1 + J'_2 )} =[J_1+J_2 \in \im \psi \phi'] (-1)^{\omega'(C', J_1'+J_2')} \sum_{A' \in \ker\psi \phi'} (-1)^{\omega'(A', J_1' +J_2')} \nonumber \\
=& [J_1+J_2 \in \im \psi \phi'] [J_1'+J_2' \in \im \psi' \phi](-1)^{\lambda(\phi'(C'), \phi(C))},
\end{align}
\end{widetext}
where $C\in \mc A$ is such that $J_1'+J_2' = \psi'\phi(C)$, we use the composite BrLE rule Eq.\eqref{eq:brle1} and we use Eq. \eqref{eq:stabgauge} in the phase. So in total,
\begin{widetext}
\begin{align}
\tr \left(p^{(J_1)}_S p^{(J_1')}_{S'}p^{(J_2)}_S p^{(J_2')}_{S'} \right) \propto  [J_1+J_2 \in \im \psi \phi'] [J_1'+J_2' \in \im \psi' \phi] [J_2 \oplus J_2' \in \im \Psi] (-1)^{\lambda(\phi'(C'), \phi(C))},
\end{align}
\end{widetext}
where we have again used Eq. \eqref{irrepID} and the BrLE rules for the subsystem gauge structure. We can include this in Eq. \eqref{Gdef}, where it is best to shift the sum $J_1 \to J_1 + J_2$, where the energy functional becomes
\begin{align}\label{Jdiff}
\|J_1\|-\|J_2\| \to& \|J_1 + J_2\| - \| J_2\| \nonumber \\
=& \|J_1\| - 2 J_1 \cdot J_2.
\end{align}
$J_1\cdot J_2 $ represents the usual real-number dot product between the vectors $J_1$ and $J_2$ as lifted to the real numbers. So the Green functions for $J_1' +J_2'= \psi' \phi(A) \in \im \psi'\phi$ and $J'_2= \psi'(f)$ for some $f\in \mc F$ can be written as

\begin{widetext}
\begin{align}\label{eq:Green}
G^2_{(S, S')}(t; J'_1, J'_2) \propto \sum_{J_1 \in \im \psi \phi'} (-1)^{\omega(A, J_1)} \exp\left( -i t\|J_1\|\right)  \sum_{J_2 \in \psi[\ker\psi']} \exp\left(i 2 t J_1 \cdot (J_2 + \psi(f))\right),
\end{align}
\end{widetext}
where we have again applied the discussion around the connection between the subsystem and composite gauge structures. Note if we take $J_1'=J_2'=0$, this reduces to Eq. \eqref{eq:0quench} as expected. 

Eq. \eqref{eq:Green} is generic for any quench between two stabilizer codes. Once we apply this to a CSS quench i.e $\phi' = \phi_X$, we can first simplify the second sum to $\psi[\ker \psi_X] =\im \psi \phi_X$ so that both dummy indices are over the same subspace. We also recognize that for $J^X_2= \psi_X(f)$, we can take $f \in \mc F_Z$. Thus $\psi(f)$ is an excitation pattern for the X-type operators of $S$. One the other hand, $J_1, J_2$ is contained in $\im \psi\phi_X$ and thus represent excitation patterns in the Z-type operators of $S$. So in the real-number dot product, $\psi(f)$ and $J_1, J_2$ have no overlap implying \break $J_1 \cdot (J_2 + \psi(f))= J_1 \cdot J_2$. Thus we can reorder the two sums and reverse Eq. \eqref{Jdiff} to obtain Eq. \eqref{eq:CSSgreen}. We then apply the identity of Eq. \eqref{irrepID} where the additional sign shifts the phase in the identity from $C\in \ker \psi_X \phi$ to $ A+C$ and we have Eq. \eqref{eq:ZofA}.

\bibliography{citations}

\end{document}